\documentstyle[11pt,newpasp,twoside,epsf]{article}
\def\etal{{\it et\thinspace al.}\ }
\def\eion{{(e~+~ion)}\ }
\def\en{{$n$}\ }

\def\edcomment#1{\iffalse\marginpar{\raggedright\sl#1\/}\else\relax\fi}
\reversemarginpar

\begin{document}
\title{The Iron Project and the RmaX Project}
  \author{Anil K. Pradhan}
\affil{Department of Astronomy, The Ohio State University, Columbus,
OH 43210, USA
({\it www.astronomy.mps.ohio-state.edu/$\sim$pradhan})}

\begin{abstract}
   Ongoing activities under an international collaboration of atomic
physicists and astrophysicists under the Iron Project and the RmaX Project, 
with applications to X-ray astronomy, are briefly described.
\end{abstract}

\section{Introduction}

 The Iron Project (IP; Hummer \etal
1993) is an extension of the erstwhile Opacity Project (OP; Seaton \etal
1994), devoted primarily to collisional and radiative processes of the
Iron-peak elements. The RmaX Project is a part of the IP aimed at
X-ray astronomy. The IP/RmaX work deals with highly charged ions and
inner-shell processes.

 To date 55 publications on {\bf Atomic Data From the Iron Project} have
appeared in Astronomy and Astrophysics. More details are on the IP
website www.usm.uni-muenchen.de/people/ip/iron-project.html, or the
author's website above. Additional details are provided in reviews in
this volume by Palmeri and Mendoza on the OP/IP database TIPTOPBASE, 
and by Nahar on "New Radiative Data" not yet generally available. 
The IP/RmaX collaboration consists of about 20 members from Canada,
France, Germany, UK, US, and Venezuela.
 Some RmaX publications are also reported in the Journal of Physics B:
Atomic, Molecular, and Optical Physics.

\section{Methodology} The IP/RmaX calculations, like the OP, are carried out
using the R-matrix method, based on the close-coupling approximation
from atomic collision theory (Burke
and Robb 1975; Seaton 1987). Unlike the OP radiative calculations that
were in LS coupling, the IP/Rmax
calculations generally take account of fine structure and some relativistic 
effects using the Breit-Pauli R-matrix method (BPRM; Berrington \etal 1995).

\section{Radiative and Collisional Calculations}
 One of the primary activities under the IP has been collisional
calculations for all Fe ions (see references in the IP series), and
radiative data for a many Fe and other ions. 

 Most recent collisional work has been on highly charged ions from H-like to 
Ne-like sequences, and K- and L-shell radiative transitions. In the
following subsections we
exemplify the nature of the IP/RmaX work, on various atomic processes
and using the same approximation (BPRM), for
the important ion Ne-like Fe~XVII that
gives rise to a number of well known X-ray lines (see the Grotrian
diagram in Chen \etal 2003).

\subsection{Electron impact excitation}
 BPRM calculations for a benchmark study of electron scattering with Fe~XVII 
showed extensive series of resonances that significantly enhance the
effective (averaged) cross sections and rate coefficients (Fig. 1, Chen and
Pradhan 2002; Chen \etal 2003). These calculations resolved longstanding
discrepancies between two sets of experimental measurements using
Electron-Beam-Ion-Traps (EBIT) at the Lawrence Livermore National Laboratory
(Brown \etal 2001) and at the National Institute for Standards and Technology 
(NIST, Laming \etal 2000). The measured and calculated cross sections
and line ratios in question are due to three prominent {\sc x}-ray
transitions labeled 3C,3D, and 3E, to the
ground level 1s$^2$2s$^2$2p$^6$~$^1$S$_0$ from excited levels:
3C ($\lambda$~15.014$\AA$)
1s$^2$2s$^2$2p$^5$[1/2]3d$_{3/2}$~$^1$P$^o_1$,
3D $\lambda$~15.265$\AA$
1s$^2$2s$^2$2p$^5$[3/2]3d$_{5/2}$~$^3$D$^o_1$, and
3E $\lambda$~15.456$\AA$:
1s$^2$2s$^2$2p$^5$[3/2]3d$_{5/2}$~$^3$P$^o_1$.
While the 3C is dipole allowed, the
3D and 3E are spin-forbidden intercombination transitions.
The so called '3s/3d' problem, also due to discrepancies between the two
sets of EBIT measurements, has also been solved using (a)
the gaussian average, and (b) the maxwellian average, over the cross
sections in the collisional-radiative model (Chen and Pradhan, in
preparation).
Chen \etal (2003) discuss the factors that affect the accuracy of 
the collision strengths for many other transitions up to \en = 4 levels 
in Fe~XVII.

\begin{figure}
\plotone{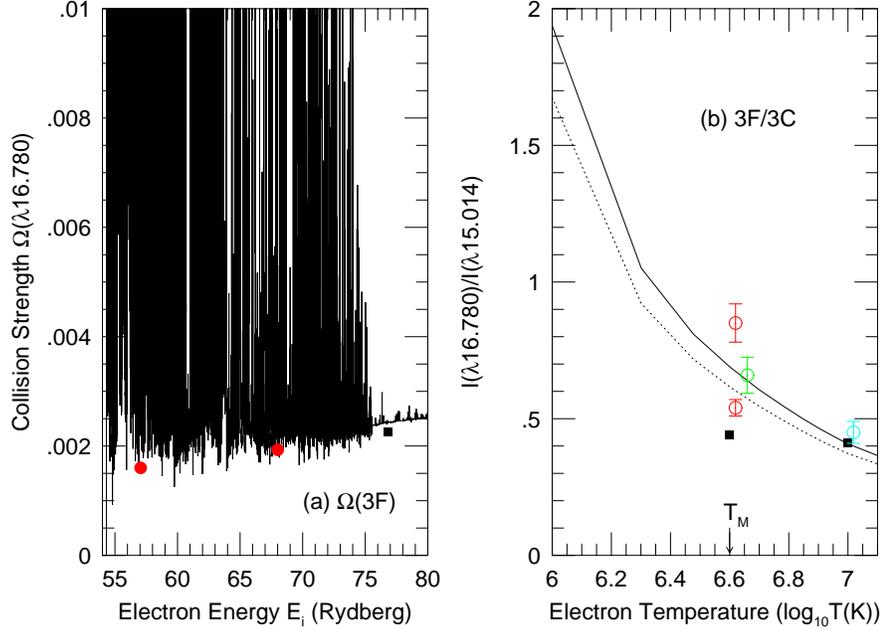}
\caption{Enhancement of collisional rates of Fe~XVII by resonances: 
(a) BPRM collision strength $\Omega$ for the forbidden 3F line
$2s^22p^53s \rightarrow 2s^22p^6, J = 1 \rightarrow 0, \lambda 
16.780 \AA$;
the filled circles and square are non-resonant DW calculations;
(b): line ratio 3F/3C vs.~T from a 89-level C-R model.
The electron densities for
solid-line and dot-line curves are 10$^{13}$ and 10$^9$ cm$^{-3}$
respectively.
The 4 open circles with error bars are observed and experimental values:
from the solar corona
at T$_m \sim$~4MK, from the corona of solar-type star Capella at $\sim$~5MK,
and from the EBIT experiment at 0.9~keV (log T = 7).
The filled squares are values using DW results. Owing to Maxwellian
averaged rate coefficients, resonance enhancement is particulary
large at low temperatures, such as in photoionized plasmas, as opposed
to higher temperatures in coronal equilibrium.}
\end{figure}

\subsection{Transition probabilities}

Relativistic BPRM transition probabilities for Fe~XVII have been calculated for
over 2.6$\times 10^4$ allowed (E1) transitions that are of dipole and
intercombination type, and about 3000 forbidden transitions that
include electric quadrupole (E2), magnetic dipole (M1), electric
octopole (E3), and magnetic quadrupole (M2) type, representing the most
detailed calculations to date for the ion (Nahar \etal 2003).

\subsection{Photoionization and Electron-Ion Recombination}

 BPRM photoionization and recombination calculations for (h$\nu$ + Fe~XVII
$\longleftrightarrow$ (e~+~Fe~XVIII) have been reported by Zhang \etal (2001),
using the unified method for \eion recombination that includes both the
radiative and the dielectronic recombination processes in an ab inito
manner. Both the level-specific and the total cross sections for the two
inverse processes are obtained. In an earlier work, Pradhan \etal (2001)
demonstrated that the theoretical unified rates agree with experimental
data from ion storage rings to within 20\%.
The unified and self-consistent approach to photoionization and \eion
recombination is reviewed by Nahar and Pradhan (2003, astro-ph/0310624).

\subsection{Opacities, inner-shell excitation, and databases}
 The review by Palmeri and Mendoza on TIPTOPBASE describes recent 
calculations on
K-shell Auger processes and the Opacity Project/Iron Project atomic and
opacities databases.

\section{Summary}

 Atomic data for a variety of processes and ions are being calculated
under the Iron/RmaX projects. The R-matrix approach is capable of
taking account of all important atomic effects, and produce data of
definitive accuracy that can be bencmharked against state-of-the-art
experiments. Self-consistent ab initio calculations for Fe~XVII are
presented as an example of large-scale data obtained for all collisional and
radiative processes in an ion using the same basic approximation (BPRM method) 
and wavefunction expansions.

\acknowledgments
 The work repored herein is partially supported by the NASA Astrophysical
Theory Program and the Space and Astrophysics Research program.

\end{document}